# Shear Thickening of Cornstarch Suspensions

Abdoulaye Fall[1,2], François Bertrand[2], Guillaume Ovarlez[2] and Daniel Bonn[1,3]

[1]*Van der Waals – Zeeman Institute, University of Amsterdam, Valckenierstraat 65, 1018XE Amsterdam, The Netherlands*
[2]*Université Paris Est, Laboratoire Navier, UMR 8205 CNRS-ENPC-IFSTTAR, 2 Allée Kepler 77420 Champs sur Marne, France*
[3]*Laboratoire de Physique Statistique, Ecole Normale Supérieure, 24 Rue Lhomond 75231Paris Cedex 05, France*

We study the rheology of cornstarch suspensions, a non – Brownian particle system that exhibits discontinuous shear thickening. Using Magnetic Resonance Imaging (MRI), the local properties of the flow are obtained by the determination of local velocity profiles and concentrations in a Couette cell. For low rotational rates, we observe shear localization characteristic of yield stress fluids. When the overall shear rate is increased, the width of the sheared region increases. The discontinuous shear thickening is found to set in at the end of this shear localization regime when all of the fluid is sheared: the existence of a non-flowing region thus seems to prevent or delay shear thickening. Macroscopic observations using different measurement geometries show that the smaller the gap of the shear cell, the lower the shear rate at which shear thickening sets in. We thus propose that the discontinuous shear thickening of cornstarch suspensions is a consequence of dilatancy: the system under flow attempts to dilate but instead undergoes a jamming transition because it is confined. This proposition is confirmed by an independent measurement of the dilation of the suspension as a function of the shear rate. It is also explains the MRI observations: when flow is localized, the non-flowing region plays the role of a "dilatancy reservoir" which allows the material to be sheared without jamming.

## I. Introduction:

Shear thickening has been observed for a wide variety of suspensions [Barnes, (1989)]. The phenomenon is frequently encountered during the processing of concentrated dispersions in various industries where it has a strong impact on energy consumption. Shear thickening can be defined as an increase in the steady-state shear viscosity $\eta$ of a fluid with the shear rate $\dot{\gamma}$ when the latter exceeds some critical value $\dot{\gamma}_c$. The detailed mechanism of shear-thickening is still under debate [see e.g. Boersma et al. (1990), Franks et al. (2000), Hoffman (1972), Macias et al. (2003), Laun et al (1992), Hoffman (1974), Foss and Brady (2000), Ackerson (1990) and Chen et al (1994). The majority of investigations of shear thickening were conducted on colloidal suspensions (Boersma et al. (1990), Franks et al. (2000), Hoffman (1972)]. Bender and Wagner (1996) and Maranzano and Wagner (2002) attribute the phenomenon to the shear-induced formation of hydrodynamic clusters – transient concentration fluctuations that are driven and sustained by the applied shear field. The viscosity rise is continuous at low volume fractions, but can also be discontinuous at higher ones [Macias et al. (2003), Laun et al (1992), O'Brien and Mackay (2000), Bertrand et al. (2002)], probably because of aggregation of hydroclusters creating a jammed network [Hoffman (1974), Foss and Brady (2000)]. In the latter case, the clustered shear thickened state may be a long–lived metastable state characterized by a large yield stress, as shown by Cates et al. (2005). In the picture of Bender and Wagner (1996), the formation of flow–induced hydroclusters results in an increased dissipation of energy and, consequently, the viscosity increases. The commonly accepted picture for shear thickening in Brownian suspensions is now the formation of shear-induced hydroclusters [see e.g. the recent summary by Wagner and Brady (2009)]. This results in large hydrodynamic stresses in rapid flowing suspensions, and there is simulation and other evidence that these are the dominant stresses in the shear-thickened regime [Bender and Wagner (1996), Phung et al. (1996)]. However, some of these simulations are limited to relatively modest volume fractions ($\varphi \leq 0.49$); experiments at much higher concentrations suggest that instead a thermodynamic mechanism may take place [O'Brien and Mackay (2000)]. In any case it seems highly unlikely that the 'static jamming' observed by Bertrand et al. (2002) is due to hydrodynamic interactions alone, since in the persisting solid phase there is no macroscopic flow to provide those interactions. Also, as far as we know, hydrodynamic models for shear thickening offer no immediate explanation of nonmonotonic regions of the flow curve [Holmes et al. (2005)]. Thus it seems clear that mechanisms other than pure hydrodynamics are at work in sufficiently dense shear-thickening suspensions [O'Brien and Mackay (2000), Bertrand et al. (2002)]. Indeed, it has already been emphasised [Ball and Melrose (1995), Melrose and Ball (1995; 2004a,b)] that deviations from pure lubrication forces can dominate the physics of any hydrodynamically clustered state [Holmes et al. (2005)].

Shear thickening is also observed in non-Brownian suspensions with larger particle sizes [Williamson and Heckert (1931); Fall et al. (2008; 2010); van der Werff and de Kruif (1989); Sellitto and Kurchan (2005) Berthier, et al (2000); Brown and Jaeger (2009; 2010)]; here the mechanisms at play are less clear. Recently, Brown and Jaeger (2009) have shown a transition between a shear thinning and a shear thickening regime where the shear thickening behavior is characterized by $\sigma \propto \dot{\gamma}^{1/n}$. For their measurements $n \approx 0.5$ far from jamming and $n$ continuously decreases towards 0 upon approaching jamming (that is, the volume fraction $\varphi \to \varphi_{max}$). However Fall et al. (2010), in a non-Brownian particle suspensions similar that those of Brown and Jaeger show that the intrinsic behavior (from local MRI measurements) shows only viscous ($n = 1$) or granular scaling (with $n = 0.5$) and that shear thickening simply corresponds to the transition between the two regimes. Note that, we call this regime granular; however it is important to realize that the scaling law with $n = 0.5$ can correspond to any inertial flow, including that of Newtonian fluids, and is not necessarily a granular scaling. The MRI data showed that in steady-state such systems are heterogeneous due to particle migration, and that consequently the macroscopic stress–strain rate relationship cannot be directly related to the local constitutive behavior and thus in particular to the shear thickening.

Here, we compare local and global measurements for what is perhaps the best – known example of a shear thickening suspension: cornstarch particles suspended in water. We show that the shear thickening can in fact be viewed as a re – entrant solid transition in this system: (i) at rest the material is solid because it has a (small) yield stress; (ii) for low shear rates, shear banding (localization) occurs, and the flowing shear band grows with increasing shear rate, the shear thus liquefies the material; (iii) shear thickening happens at the end of the localization regime, where all of the material flows, subsequently it suddenly becomes ''solid'' again. In addition, (iv) we find a pronounced dependence of the critical shear rate for the onset of shear thickening on the gap of the measurement geometry, which can be explained by the tendency of the sheared system to dilate. This is confirmed by an independent measurement of the dilation of the suspension as a function of the shear rate. It also explains the MRI observations: when flow is localized, the non-flowing region plays the role of a "dilatancy reservoir" which allows the material to be sheared without jamming.

This paper follows up on our earlier work on shear thickening of cornstarch [Fall et al., (2008)], but is much more detailed in that here we present also the MRI measurements of the concentration, more detailed measurements of the velocity profiles, plate-plate measurements, oscillation measurements and more detailed measurements of the variation of the gap of the plate-plate cell under an imposed normal stress. In order for these new data to be comprehensible we do have to repeat some of the earlier data and discussion. In this way we obtain a more complete picture of the shear thickening behavior.

## II. Materials and methods

The cornstarch particles (from Sigma Aldrich) are relatively monodisperse particles with, however, irregular shapes [Figure 1]. Suspensions are prepared by mixing the cornstarch with a 55 wt% solution of CsCl in demineralized water. The CsCl allows one to perfectly match the solvent and particle densities [Merkt et al. (2004)]. We study suspensions of volume fraction ranging between 38% and 46%, and focus here mainly on the behavior of a 44% cornstarch suspension that is representative of the rest. The effect of changing the volume fraction will be discussed in detail in Sec.III.6.

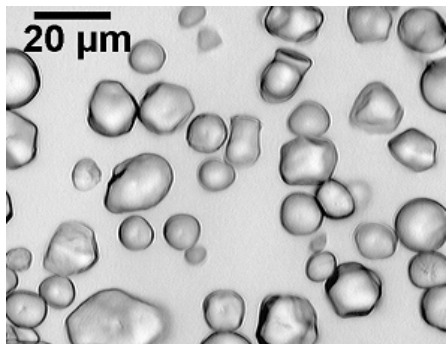

Figure 1: Micrograph of the cornstarch particles.

Experiments are carried out with a vane-in-cup (inner cylinder radius Ri = 12.5 *mm*, outer cylinder radius Re = 18.5 *mm*, height H = 45 *mm*) or parallel plate geometry on a commercial rheometer (Bohlin *C-VOR 200*) that imposes either the torque or the rotational velocity (with a torque feedback). The vane geometry is equivalent to a cylinder with a rough lateral surface which reduces wall slip [Larson (1999)]. The inside of the cup is also covered with the granular particles using double sided adhesive tape. For the parallel plate geometry, the upper plate is of 40 *mm* diameter; both plates are roughened. Velocity profiles in the flowing sample were obtained with a velocity controlled magnetic resonance imaging (MRI) rheometer from which we directly get the local velocity distribution in a Couette geometry with a gap of 1.85 cm. We investigated the stationary flows for inner cylinder rotational velocity Ω ranging between 0.2 and 10 *rpm*, corresponding to overall shear rates between 0.04 and 2.35 s$^{-1}$ [Raynaud et al. (2002), Rodts et al.

(2005), Bonn et al.(2008)]. All the measurements were done in the same laboratory with a controlled humidity of 40%; cornstarch is likely to take up some water from the atmosphere.

## III. Experimental results

### 1. Typical macroscopic behavior

Let us first present the typical behavior observed when shearing a cornstarch suspension (Figure 2)

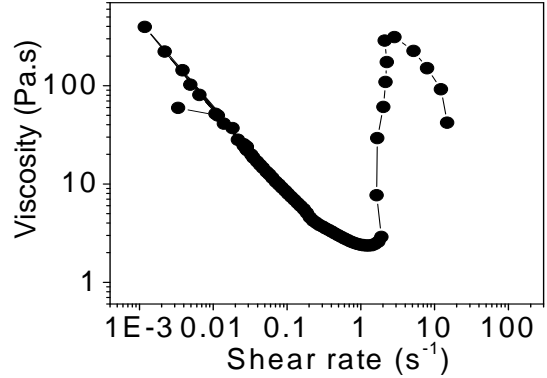

Figure 2: Apparent viscosity *vs*. shear rate when performing a step stress test in a vane in cup geometry.

The viscous properties at higher stresses were measured with step stress tests. The basic trends in $\eta(\dot{\gamma})$ are similar for all measured volume fractions: at low stresses, $\eta$ decreases with increasing applied stress, reflecting shear-thinning behavior. As the stress increases further, $\eta$ increases, reflecting shear-thickening behavior, and then reaches a plateau. This abrupt increase in viscosity observed is characteristic of "discontinuous" shear thickening. To better understand this behavior, we performed MRI measurements in a Couette geometry.

### 2. Local rheology: velocity and concentration profile measurements

In Figure 3, we plot the dimensionless velocity profiles for the steady flows of a cornstarch suspension, for various rotational velocities ranging from 0.2 to 9 rpm.

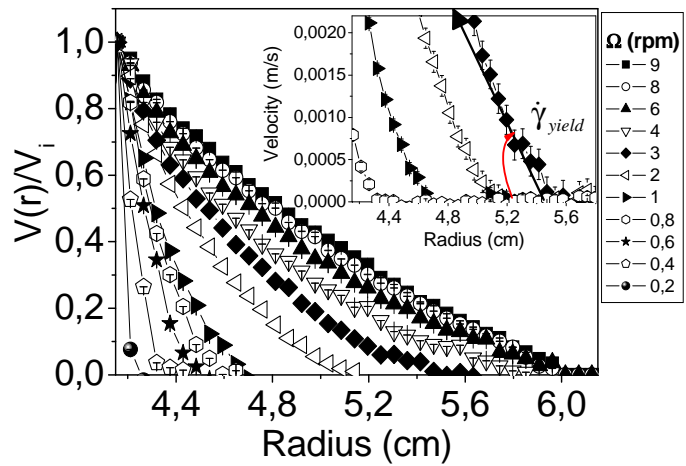

Figure 3: Dimensionless velocity profile in the gap obtained by MRI measurements. Insert: the shear rate at the interface between sheared and unsheared regions is given by the slope of the velocity profile at that point, taken of course in the moving part of the material.

The MRI measurements of the velocity profiles first, show that there is no slip to within the experimental uncertainty. The data shown is normalized on the speed of the rotating inner cylinder, and extrapolates to unity, showing that slip is negligible. Second, the flowing part of the sample occupies only a small fraction of the gap at low rotation velocities: we observe shear localization. The velocity profiles are composed of two regions: the part close to the inner cylinder is moving, and the rest is not. For the lower rotation speeds, since the part of the material that does not move is subjected to a stress, this implies that the suspension has a yield stress. The yield stress can be determined from the critical radius $r_c$ at which the flow stops: the shear stress at a given radius $r$ as a function of the applied torque $T$ and fluid height $H$ follows from momentum balance, and thus the yield stress at $r_c$ follows immediately as $\sigma_y = T/2\pi H r_c^2$. The yield stress turns out to be on the order of 0.3 Pa. Although it seems obvious that concentrated suspensions that show shear thickening also have a yield stress, we have not found literature comparing the pre-thickening flow behavior to a Bingham model as we do here, with the exception of the recent work of Brown and Jaeger (Brown and Jaeger, 2010) where the pre-thickening behavior was compared to a Herschel-Bulkley model. This is probably due to the fact that the yield stress is low. We can detect it relatively easily here because we use the MRI data.

A striking observation is that the shear rate at the interface between sheared and unsheared regions is different from zero, in contrast with what is observed in simple yield stress fluids [Bonn and Denn (2009), Moller et al. (2009), Ovarlez et al., 2008, 2010]. We indeed observe that the slope of the velocity profile at that point [Figure 3, Insert] is equal to 0.2s$^{-1}$. This implies that below $\dot{\gamma}_{yield} \approx 0.2 s^{-1}$ there is no stable flow. The existence of a critical shear rate for yield stress materials has been discussed previously in detail, and requires that the system be (slightly) thixotropic. The critical shear rate associated with the yielding of thixotropic materials has been discussed in detail elsewhere [Moller et al. (2006), (2008), (2009)]. In our case, this is likely to be due to competition between slight sedimentation or creaming, i.e., a density matching that is not perfect and shear-induced resuspension [Coussot et al., (2002a, b); Fall et al. (2009)]. Upon increasing the rotation rate, a larger part of the fluid is sheared, and for the highest rotation speeds the sheared region occupies the entire gap. We are unable to go to higher rotation rates in the MRI since the shear thickening sets in immediately when the shear band occupies the entire gap of the Couette cell, and when it does the motor of the rheometer is no longer sufficiently strong to rotate the inner cylinder: shear thickening is observed as an abrupt increase of the measured torque on the rotation axis. We therefore see the onset of shear thickening at the first shear rate for which all of the material is sheared; we will show below that this is likely to happen because the presence of a non-flowing region delays and attenuates the shear thickening.

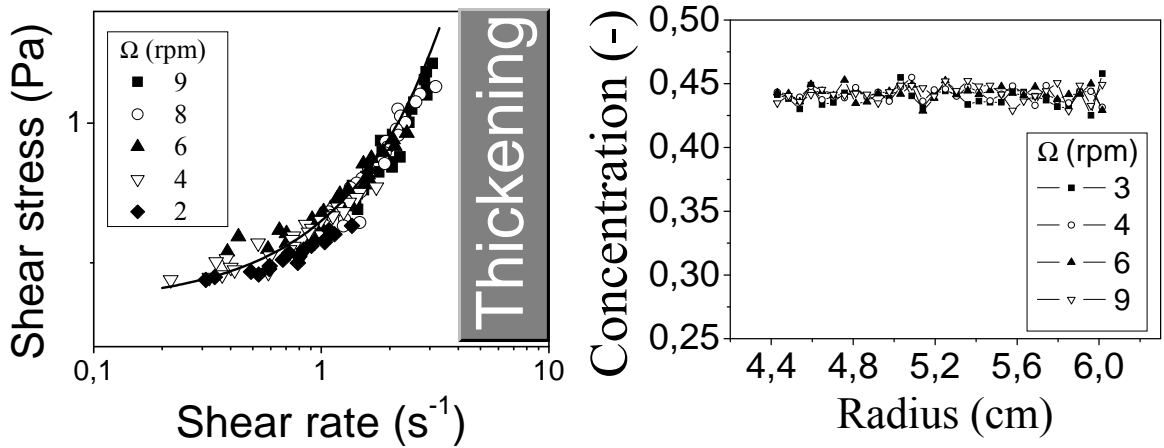

Figure 4: (a) Local shear stress as a function the local shear rate. The line is a fit to the Bingham model: $\sigma = \sigma_y + k\dot{\gamma}$ with $\sigma_y \approx 0.35 Pa$, $k = 0.30 Pa.s$. b) Local concentration profiles in the Couette gap geometry for a cornstarch suspension sheared at various rotational velocities ranging from 3 to 9 rpm.

From the velocity profiles $v(r)$, we can determine the constitutive behavior. In the flowing part, the local shear rate within the Couette gap can be given as $\dot{\gamma}(r) = \left(\dfrac{\partial v}{\partial r}\right) - \dfrac{v}{r}$. By combining $\dot{\gamma}(r)$ with the local shear stress $\sigma(r) = T/2\pi H r^2$ at each radial position $r$, for various rotational velocities $\Omega$, one obtains the constitutive equation $\sigma(\dot{\gamma})$ of the fluid. The locally measured constitutive equation of the cornstarch suspension is shown in Figure 4 (a). It is consistent with the observation that $\sigma_y \approx 0.3 Pa$ and $\dot{\gamma}_{yield} \approx 0.2 s^{-1}$.

The MRI also allows us to measure the particle concentration; to within the experimental uncertainty of ±0.2% in volume fraction the particle concentration is homogeneous throughout the gap. Thus, during the flow and at the onset of shear thickening, the suspension remains homogeneous within the gap. As a result, no migration of particles is observed to the limit of the transition to the shear thickening regime. This is perhaps surprising in the light of recent measurements of shear thickening in suspensions of spherical particles, where shear thickening is always accompanied by particle migration (Fall et al, 2010). In Couette flows, the consequence of migration is an excess of particles near the outer cylinder [Leighton and Acrivos (1987a, b), Ovarlez et al. (2006), Huang and Bonn (2007)]. However our measurements do not completely rule out particle migration; we estimate, from the MRI data that the maximum gradient in particle concentration, if any, is around 0.1%. This is indeed very small when compared to migration for suspensions of spherical particles, where gradients of several percents are observed in the same MRI Couette cell.

The critical shear rate is due to the existence of a yield stress and is in principle decoupled from the onset of shear thickening. However, in the Couette geometry used for the MRI experiments, the existence of a yield stress makes that part of the material flows, and another part does not because the stress it is subjected to is smaller than the yield stress. As shown by the plate-plate presented below, the existence of a non-flowing region influences the onset of shear thickening, and so the critical shear rate is indirectly coupled to the shear thickening phenomenon. To the contrary, for the critical stress both the dilation and the measurements as a function of volume fraction show that the onset of shear thickening is directly dependent on the stress.

The emergence of shear thickening at the end of the localization regime then suggests that the non-flowing part plays an important role in the shear thickening. To assess what this role is, classical rheology measurements were conducted.

## 3. Role of a dead zone

In order to investigate in detail the influence of the non-flowing region in the Couette cell on the observed shear thickening, parallel plate geometry is used. This geometry has the additional advantage that there is no reservoir of particles present, as is the case of the Couette. If we need to, a 'non-flowing' region can be created in this geometry by simply leaving a few milliliters of paste around the gap and in contact with the sample between the plates [Figure 5]. In addition this geometry allows us to measure the normal stresses.

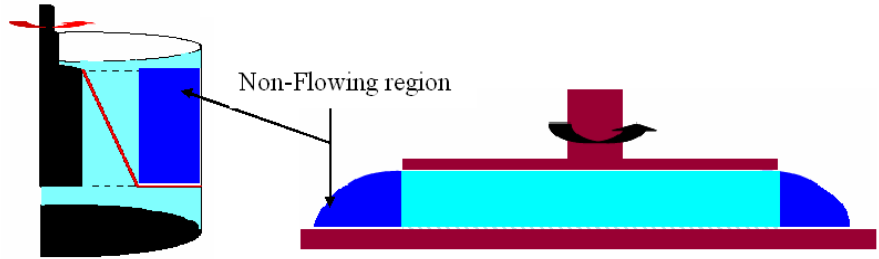

Figure 5: example of a 'non–flowing' region or 'dead zone' in parallel plate geometry.

The rheometer measures a torque $T$ and a rotation rate $\omega$, which are related to the stress and shear rate at the edge $r = R$ of the sample by $\sigma = 3T/2\pi r^3$ and $\dot{\gamma} = 2\pi r\omega/b$, with $R$ the plate radius and $b$ the gap size. In this case the shear rate being up and shear thickening is favored. The viscosity and normal stresses are measured with a gap size of 0.8 *mm*.

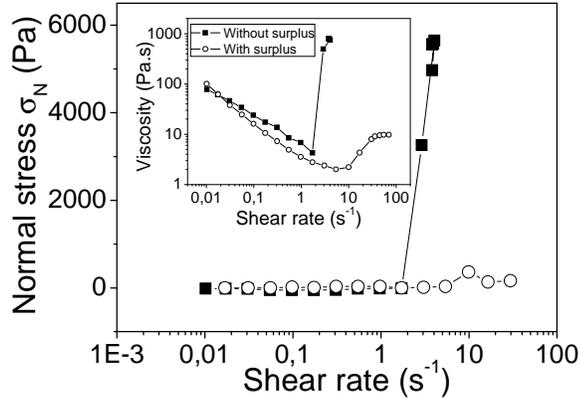

Figure 6: Role of surplus of paste on the shear thickening transition: Evolution of normal stress and viscosity as a function of the applied shear rate.

Figure 6 shows the evolution of viscosity and normal stress with the applied shear rate. For low shear rates, a typical behavior $\sigma \propto 1/\dot{\gamma}$ of a shear thinning fluid is observed. At a certain shear rate, a very abrupt increase in viscosity is observed. However, this abrupt increase in viscosity is only observed when the surplus of paste around the plates is carefully removed. If a few milliliters of suspension are left on the bottom plate in contact with the paste between the two plates, the shear thickening is much attenuated: there is no abrupt increase in viscosity. Defining the critical shear rate as the first shear rate for which the apparent viscosity goes up, the surplus of paste increases the critical shear rate very significantly.

We also observe that the shear thickening is accompanied by the emergence of large normal stresses. For low shear rates, the normal stresses are very small. However, from the critical shear rate on, grows and becomes very important. The critical shear rate for which normal stresses appear is very similar to that for which the viscosity increases. We note also that if a surplus of paste is left around, the shear thickening is accompanied by much lower normal stresses.

These results provide a possible explanation for the MRI experiments in which shear thickening only happens when all of the material flows in the Couette geometry. When only part of the material flows, the dead zone plays a role analogous to the surplus in the parallel plate experiments, which delays and weakens the shear thickening, as is observed here in the parallel plate geometry. At the end of the shear localization regime, the Couette system is suddenly analogous to the parallel plate geometry without surplus; the critical shear rate is thus suddenly lowered and the material jams. In conclusion, it seems evident that, indeed, the presence of a non-flowing region delays and attenuates the shear thickening. We can note that these findings are consistent with a picture that hydroclusters formation leads to discontinuous shear thickening and jamming when the hydroclusters percolate the structure [Wagner and Brady, (2009); Cates et al (2005)].

## 4. Dilation effect

These data suggest that, in both measurement geometries, the dead zone plays the role of a "reservoir" of dilation which helps the material to flow without jamming. Indeed, the principal information obtained from the normal stress measurement is their on-off behavior, which is quantitatively linked with the onset shear rate of shear thickening, as was verified by studying different volume fractions [Fall et al. (2008)]. The normal stresses are reminiscent of the shear induced dilatancy of dry granular matter: when sheared, it will dilate in the normal direction of the velocity gradient. Dilatancy is a direct consequence of collisions between the grains: to accommodate the flow, the grains have to roll over each other in the gradient direction, and hence the material will tend to dilate in this direction. However, in our system without a surplus, the grains are confined, both between the plates and in the solvent. The latter provides a confining pressure that is mainly due to the surface tension of the solvent, making it impossible to remove grains from the suspension. As suggested by Cates et al (2005) and Fall et al. (2008), the confinement pressure associated with this should be on the order of the surface tension over the grain size, $P_c = \gamma/R \approx 7000 Pa$. This confinement pressure is in the same order of magnitude as the typical normal stresses measured near the onset of shear thickening [Figure 6]. In addition, this gives a maximal dilation that is on the order of 1 particle diameter ($\approx 20 \mu m$); compared to the radius of the parallel plate cell this gives a maximum dilation of about 0.1%, too small to be detected by our MRI density measurements.

In dilatancy [Reynolds (1885)], the volume of a collection of particles must increase upon shearing to enable flow. This has been suggested as a possible mechanism for jamming in concentrated colloidal suspensions [Cates et al. (2005)]. Dilation within a fixed volume of suspending liquid involves the formation of force transmitting 'hydroclusters', whose growth eventually causes particles to encounter the air-liquid interface. This generates large capillary forces at the free surface, which can then balance the normal inter-particle forces and resist further motion. The particles can thus form spanning hydroclusters in close contact, jamming the sample. This may then fracture into millimetre-scale 'granules' [Cates et al. (2005)]. Jamming of colloids is also seen in pipe and channel flows [Isa et al. (2009); Haw (2004)]; here free surfaces are not present. As the upper plate is retracted, a filament forms which narrows and eventually breaks. An elongational flow, in contrast to more conventional shear and pipe geometries, therefore implies an increase of the interfacial area during flow. Although purely elongational flow can be achieved by exponential plate separation, a constant separation speed is closer to fiber-spinning and other industrial processes. In these, a purely tensile loading evokes a mixed flow combining elongation and shear. Recent studies [Bischoff White et al. (2010)] have demonstrated that extensional rheometry can successfully be performed on colloidal suspensions. Bischoff White et al measured the tensile stresses of a ($\varphi \sim 0.355$) cornstarch solution. They observed a flowable filament at low extension rates but beyond this found a transition to brittle fracture. The interactions in this system are poorly characterized but clearly attractive (see Figure 1 of [Bischoff White et al. (2010)]), presumably due to strong van der Waals forces. Such interactions could strongly influence the flow behavior as they do in strongly aggregating colloids at lower densities [Bischoff White et al. (2010)].

We further investigate the role of dilation by doing oscillatory rheology (Figure 7). We find that, the material shows a linear elastic behavior $G' \approx 10\, G''$ at low shear stresses, i.e. it behaves as a "solid" material. Increasing the stress, a "solid-liquid" transition is found characterized by a yield stress at which $G' \approx G''$; then the suspension begins to flow and $G' < G''$. These observations are consistent with the behavior observed in steady shear (see above). Increasing the applied stress even more, the 'liquid' regime quickly ends by the shear thickening of the system (characterized by an abrupt increase of G' and G''), for a critical shear strain $\gamma_c \approx 1$

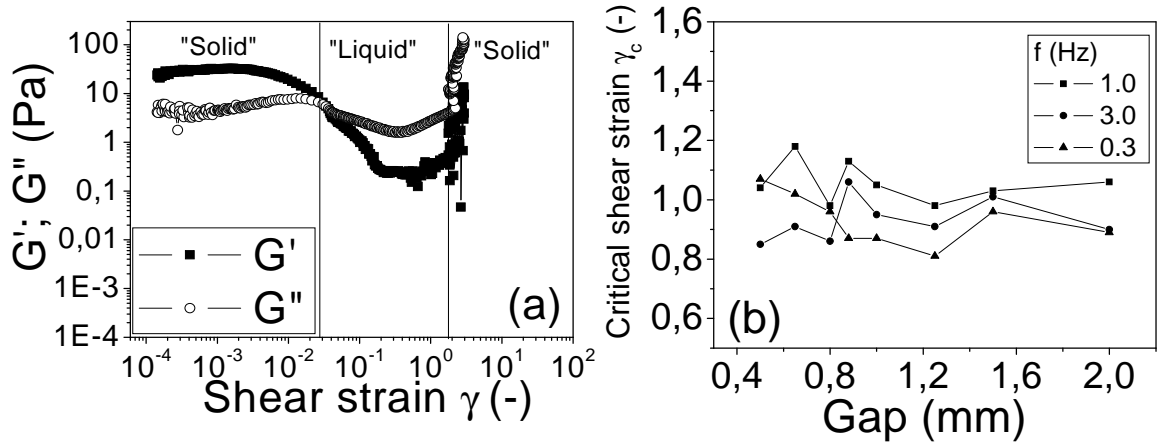

Figure 7: (a) Elastic modulus (G') and loss modulus (G'') in the 44% volume fraction suspension as a function of shear strain for an imposed shear stress (0.001 to 100 Pa) at 1 Hz in a vane geometry. (b) Critical strain vs. gap and frequency

It is clear from Figure 7 that the nonlinearity between stress amplitude and strain amplitude becomes more and more significant as the stress amplitude increases. Moreover, the appearance of nonlinear behavior is accompanied by non-sinusoidal responses in the oscillatory shear experiments. As the higher harmonic signals emerge, the material functions such as G' and G'' lose their original physical meaning. Considerable efforts [Yu et al. 2009; Cho et al. 2005; Ewoldt et al. 2008; Klein et al. 2007] have been made towards obtaining useful and desirable material information from such LAOS (Large Amplitude Oscillatory) experiments. For our purposes, the nonlinear oscillatory experiments are useful for characterizing the onset of shear thickening as well for determining the time scales required to generate the shear thickening response. Indeed, in the oscillatory experiments, the first natural interpretation of the observation of shear thickening would be that the shear rate $\dot{\gamma} = 2\pi \gamma f$ applied during the oscillations is equal to the critical shear rate $\dot{\gamma}_c$ observed in the continuous shear experiments when $\gamma \approx \gamma_c$. With a frequency $f$ =1Hz, the critical oscillatory shear rate at the onset of thickening is $\approx 6.5$ s$^{-1}$, a value 3 times higher than the critical shear rate observed during continuous shear. To understand this difference, we have performed the same experiments for several frequencies. We observe, in Figure 7(b), that the critical shear strain $\gamma_c \approx 1$ at the onset of thickening is constant; it is also independent of the gap sized, in contrast with $\dot{\gamma}_c$ (see below). The conclusion is that $\gamma_c \approx 1$ is indeed the relevant physical quantity in the oscillatory experiments. This observation means that, even for high shear rates, thickening cannot take place if a two neighboring grains did not experience a relative motion of order of particle diameter; this is consistent with the major role of dilation evidenced above.

## 5. Confinement effect

It is therefore tempting to see whether the shear thickening phenomenon itself can be attributed entirely to the confinement: if the cornstarch is confined in such a way that the grains cannot roll over each other, this could in principle lead to an abrupt jamming of the system.

Experimentally, instead of setting the gap size in the rheometer for a given experiment, we can impose the normal stress and vary the gap size in order to reach a target value of the normal stress. If this is done for different shear rates, and the target value for the normal stress is taken to be zero, we can obtain the dependence of the gap variation on shear rate. A typical measurement is shown in Figure 8 (a), where we impose a constant shear rate and measure the gap and viscosity as a function of time. This shear rate and initial gap combination are beyond the shear-thickening transition in shear rate, and thus the viscosity (shear stress) starts to strongly increase, as do the normal stresses. The latter then leads, through a feedback loop, to an increase in the gap, allowing the system to dilate, until the shear thickening disappears altogether: the shear stress is back to low values (less than 10 Pa). This unambiguously demonstrates that the shear thickening is a dilation effect, and that taking away a confining factor makes the thickening disappear altogether.

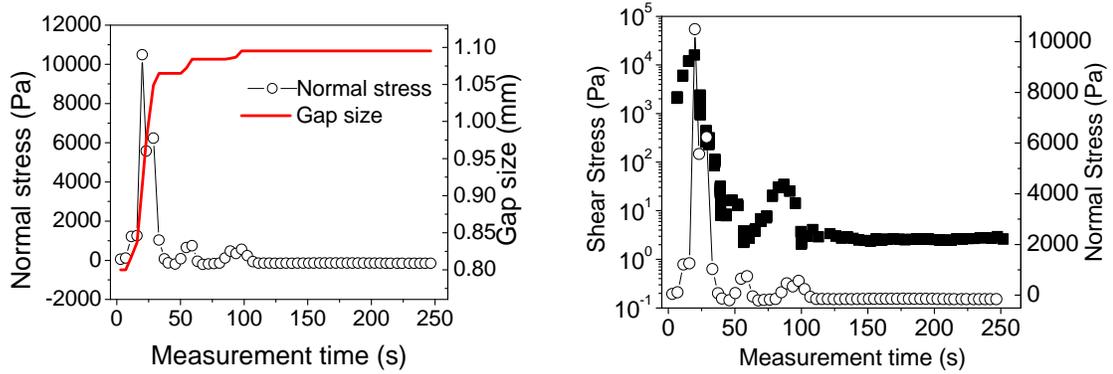

Figure 8: Time evolution for a $\dot{\gamma} = 1.6 s^{-1}$ applied shear rate of: (a) the gap size and the normal stress; (b) Normal (circles) and shear stresses (squares). The rheometer is set to change the rotation rate of the tool through a feedback loop to make sure the shear rate is constant, even though the gap is changing.

More quantitatively, repeating this experiment for different shear rates [Figure 8(b)], one can obtain the gap change as a function of the shear rate that allows the suspension to flow freely, i.e., without developing normal stresses due to particle collisions. The linear evolution of $\Delta h$ with the shear rate $\Delta h = \beta \dot{\gamma}_c$ with $\beta \approx (0.273 \pm 0.013)$ *mm.s*.

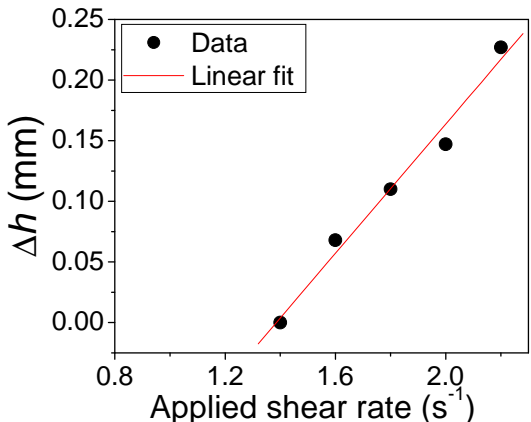

Figure 9: Variation of the gap according to the shear rate.

This can be compared to the parallel-plate experiments in which the gap was systematically varied. Figure 10(a) shows the measured apparent viscosity as a function of shear rate for different gaps. For low shear rate, again a shear thinning behavior is observed. At the critical shear rate, a very abrupt increase in viscosity is observed; the main point here is that this critical shear rate increases with increasing gap. Comparison between parallel plate, cone and plate, and Couette cells showed identical critical shear rates to within the experimental uncertainty showing that the shear rate gradient present in our parallel plate geometry does not strongly affect our results.

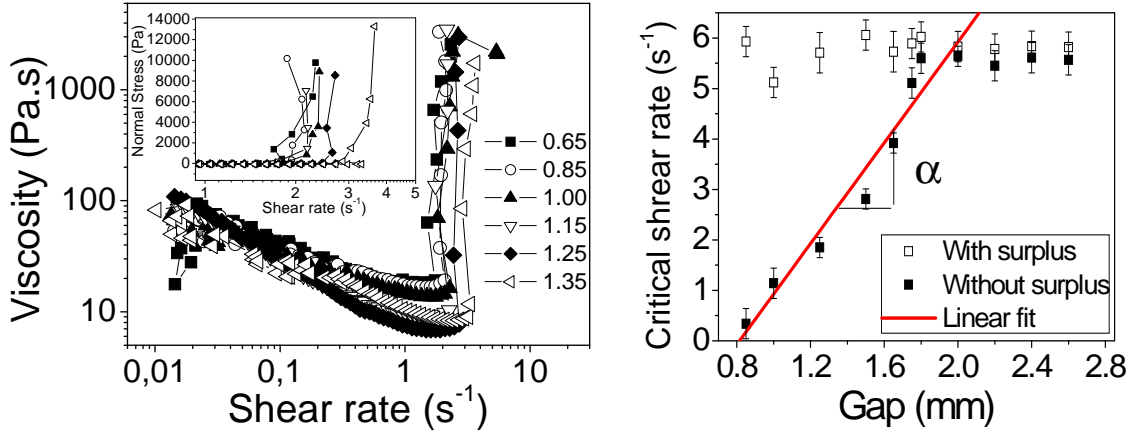

Figure 10: (a) Apparent viscosity and normal stress as a function of shear rate for different gaps in mm. Measurements were made with a parallel plate rheometer (Bohlin C-VOR 200) with radius $R = 20$ *mm*. (b) Evolution of the critical shear rate a function of the gap. The error bars correspond to the uncertainty (reproducibility) of the experiments.

We then define the critical shear rate as the shear rate for which both the apparent viscosity increases and for which non-zero normal stresses are first observed: both values are very similar and, more importantly, increase linearly with the gap [Figure 10]. Again, if a few milliliters of suspension are left on the bottom plate in contact with the paste between the two plates, the critical shear rate strongly increases and becomes roughly independent of the gap size [Figure 10].

The critical shear rate with a surplus of paste present is, in addition, the same as that found in the large gap Couette cell, in which there is also a reservoir of particles present in the non-flowing region. The flow curve of Figure 3 shows that in the MRI experiments the critical shear rate is ~ 4 s$^{-1}$; as soon as this shear rate is exceeded, the system shear thickens. We therefore conclude that the critical shear rate for thickening obeys $\dot{\gamma}_{c_M} = \dot{\gamma}_{c_I} - \alpha h$ for $h < h_c$ and constant above; here *h* is the gap size, $\alpha = (4.95 \pm 0.44)$ (*mm*.s)$^{-1}$, $\dot{\gamma}_{c_M}$ the critical shear rate when the suspension is sufficiently confined and $\dot{\gamma}_{cI}$ the intrinsic critical shear rate.

Moreover, this linear evolution of a critical shear rate with the gap size presents a striking similarity with the dilation results shown in Figure 9. Indeed, we can observe that the value of $\alpha = 4.95$ is roughly consistent with the value found in Figure 9 ($\beta^{-1} = 3.66$), providing a quantitative check that indeed the dilatancy is responsible for the shear thickening.

## 6. Shear thickening as a Viscous/Granular transition

Another parameter that has a large effect on the critical shear rate is the volume fraction. At low volume fractions: $\varphi < 0.4$, shear thickening is either less dramatic or absent [van der Werff and de Kruif (1989)]. In Figure 11, we show the typical evolution of the viscosity as a function the shear rate of cornstarch suspension with different concentrations.

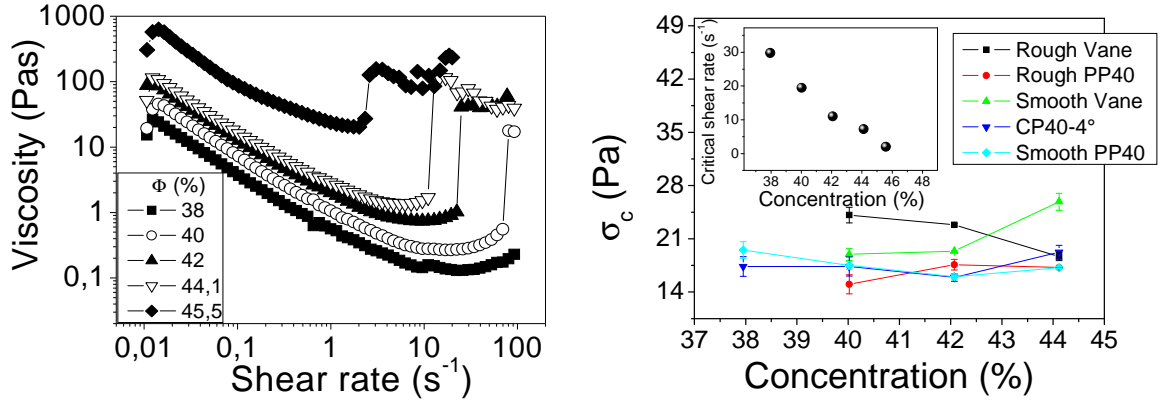

Figure 11: Viscosity as a function a shear rate for different concentrations. (b) Critical shear stress of shear thickening vs. concentration in different measurement geometries. Inset: Critical shear rate of shear thickening vs. concentration.

We found that the critical shear rate $\dot{\gamma}_c(\varphi)$ decreases roughly linearly with increasing concentration and secondly that the critical shear stress of shear thickening remains roughly constant at $\sigma_c(\varphi) \approx 20 Pa$ consistent with the work of [Fall et al. (2010)]. In this picture, for low stresses, the viscosity of the interstitial fluid lubricates the contacts and one recovers a viscous behavior; this agrees with the observed Bingham behavior provided the stress on the system is much larger than the yield stress so that the latter can be neglected. However for higher stresses, a Bagnold-type dry granular rheology would be expected. The hallmark of this behavior is that the stress in the Bagnold regime, i.e., beyond shear thickening scales inertially [Fall et al. (2010), Lemaître et al. (2009), Mills and Snabre (2009)], implying $\sigma \propto \gamma^2$. However, for the cornstarch, this cannot be verified directly because the system beyond shear thickening has a very high viscosity making measurement over a large range of shear rates impossible. In addition, for the highest shear rates one observes instabilities in both the parallel plate cell and the Couette cell that are probably due to the normal stresses, and that make that the fluid is expelled from the gap.

However remaining around the onset shear rate, we can verify that in the second, shear thickening regime, the system behaves similarly to a dry granular material, which is exactly what is at the basis of the lubricated to inertial transition. If the system is a frictional one (as a dry granular material should be), due to the steric interaction between the particles, the local shear stress $\sigma$ induces a local normal stress $\sigma = \mu \sigma_n$ where μ is the macroscopic friction coefficient.

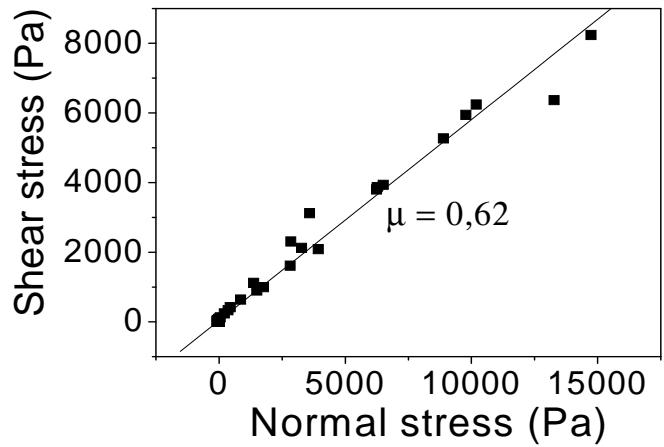

Figure 12: Proportionality between normal stress and shear stress in the shear thickening regime for the 44% volume fraction suspension

Figure 12 shows the evolution of shear stress as a function a normal stress for different gap sizes of parallel plate geometry. We found that the normal stresses change linearly with the shear stresses. This linearity can be used to define the macroscopic friction coefficient of the suspension; we find a value 0.62±0.01: this value is similar to that found in dry granular materials [da Cruz et al, (2005)] and other shear-thickening fluids [Lootens et al. (2003, 2005); Brown and Jaeger, (2011)]. This suggests that in the shear thickening regime, the interstitial fluid plays no more role as lubrication forces become negligible: contacts behave like "dry" contacts. In this case, shear thickening can therefore again be considered as a transition from lubricated regime to a frictional regime under flow. The shear thickening transition is then due to direct contacts between particles induced by the shear.

## 7. Shear thickening as a re-entrant jamming transition

In our situation, the yield stress is smaller than the critical stress and consequently the sample yields and flows before thickening. If the critical stress is the smallest one, the thickening behavior may even disappear altogether [Gopalakrishnan and Zukoski, (2004), Brown and Jaeger (2009)]. Our data show that for the cornstarch suspensions there are two critical stresses for which the apparent viscosity becomes infinite: first, upon approaching the yield stress from above, the viscosity diverges, in agreement with the MRI observations that the flow behavior is close to that of a Bingham fluid. Second, at the critical stress for thickening, an almost discontinuous jump of the viscosity is observed. These observations suggest a "solid–liquid–solid" transition. This is in addition qualitatively the same picture as that obtained from by the oscillatory shear experiments (Figure 7).

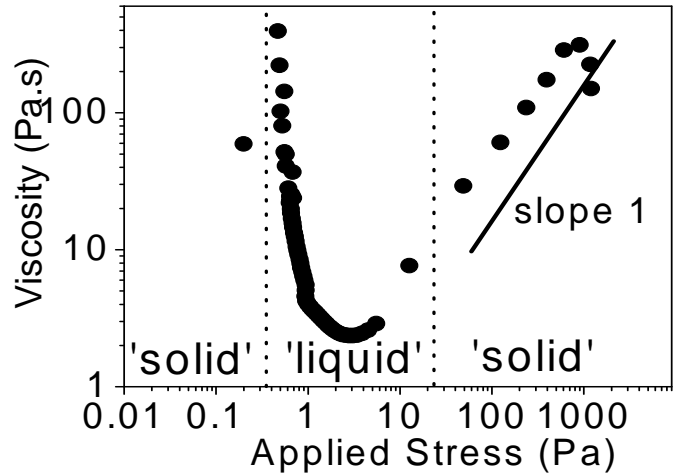

Figure 13: Solid–liquid–solid transition; apparent viscosity *vs*. stress when performing a step stress test.

This is a typical example of discontinuous shear thickening, in which the region with positive slope of $\sigma(\dot{\gamma})$ occurs in a stress range that is nearly independent of packing fraction as mentioned above in Figure 11. This slope increases with packing fraction, approaching $\eta \propto \sigma$ corresponding to a discontinuous stress/shear-rate relation. These results are similar to theory that suggests that shear thickening is due to a re-entrant jamming transition [Sellito and Kurchan (2005)]. It has been suggested for glassy systems that applying a shear is equivalent to increasing the effective temperature with which the system attempts to overcome energy barriers [Berthier et al. (2000)]. If now a system has a re-entrant ''solid'' transition as a function of temperature, the ''solid'' phase may also be induced by the shear, leading to shear thickening [Sellito and Kurchan (2005); Fall et al. (2008)], as is also observed here (Figure 13). Indeed, the sample is "solid" in the sense that G'>>G'' (Figure 7) and the torque becomes too large for the MRI rheometer to turn. In the convential rheology, there are instabilities due to the large elasticity that expel the starch from the gap of the measurement geometry, making determination of an apparent viscosity difficult.

## IV. Conclusion

We have studied the flow behavior of dense suspensions of non-colloidal particles, cornstarch particles in water, by coupling local velocity and concentration measurements through MRI techniques, and macroscopic rheometric experiments in Couette and parallel plate geometries. The MRI data reveals that the flow exhibits shear-banding at low rotation velocities of the inner cylinder. The MRI also showed that the material is homogeneous and no migration is observed, contrary to what happens for non-Brownian suspensions of spherical particles [Fall et al., (2010)].

Classical rheology then shows that the critical shear rate for the onset of shear thickening depends on the confinement: it evolves linearly with the gap size. This linear evolution is a direct consequence of the dilatancy observed in this suspension under shear. Indeed, from dilation measurements, we have shown that the application of a shear rate higher than the critical shear rate for thickening directly leads to a dilation of the suspension. Conversely, taking away the confinement attenuates the shear thickening behavior, or even makes it disappear altogether. For the MRI observations this implies that when flow is localized, the nonflowing region plays the role of a "dilatancy reservoir" which allows the material to be sheared without undergoing a jamming transition.

This convincingly shows that the dilation is at the origin of the discontinuous thickening. In addition, in oscillatory experiments we found a critical shear strain for shear thickening that appears independent of the frequency. The value of the critical shear strain ($\gamma_c \approx 1$) also supports the idea that discontinuous shear thickening is a direct consequence of shear induced dilatancy that is hindered by the confinement.

The critical shear rate associated with the yield stress is due to the existence of a yield stress and is in principle decoupled from the onset of shear thickening. However, in the Couette geometry used for the MRI experiments, the existence of a yield stress makes that part of the material flows, and another part does not because the stress it is subjected to is smaller than the yield stress. As also shown by the plate-plate measurements (Figure 6), the existence of a non-flowing region influences the onset of shear thickening, and so the critical shear rate is indirectly coupled to the shear thickening phenomenon. However both the dilation and the measurements as a function of volume fraction show that the critical stress for the onset of shear thickening is directly coupled to the onset behavior.